\documentclass[prl,twocolumn,floatix,amssymb,showpacs,amsmath,superscriptaddress]{revtex4-1}
\usepackage{graphicx}
\usepackage{epsfig,color}
\usepackage{amsmath,amssymb}
\begin{document}

\newcommand{\ket}[1]{\ensuremath{\left|{#1}\right\rangle}}
\newcommand{\bra}[1]{\ensuremath{\left\langle{#1}\right|}}
\newcommand{\quadr}[1]{\ensuremath{{\not}{#1}}}
\newcommand{\quadrd}[0]{\ensuremath{{\not}{\partial}}}
\newcommand{\slpar}{\partial\!\!\!/}
\newcommand{\gtrescero}{\gamma_{(3)}^0}
\newcommand{\gtresuno}{\gamma_{(3)}^1}
\newcommand{\gtresi}{\gamma_{(3)}^i}

\title{Quantum Simulation of Quantum Field Theories in Trapped Ions}

\date{\today}

\author{J. Casanova}
\affiliation{Departamento de Qu\'{\i}mica F\'{\i}sica, Universidad del Pa\'{\i}s Vasco -- Euskal Herriko Unibertsitatea, Apdo.\ 644, 48080 Bilbao, Spain}
\author{L. Lamata}
\affiliation{Departamento de Qu\'{\i}mica F\'{\i}sica, Universidad
del Pa\'{\i}s Vasco -- Euskal Herriko Unibertsitatea, Apdo.\ 644,
48080 Bilbao, Spain}
\author{I. L. Egusquiza}
\affiliation{Departamento de F\'{\i}sica Te\'{o}rica, Universidad del Pa\'{\i}s Vasco -- Euskal Herriko Unibertsitatea, Apdo.\ 644, 48080 Bilbao, Spain}
\author{R.~Gerritsma}
\affiliation{Institut f\"ur Quantenoptik und Quanteninformation, \"Osterreichische Akademie der Wissenschaften, Otto-Hittmair-Platz 1, A-6020 Innsbruck, Austria}
\affiliation{Institut f\"ur Experimentalphysik, Universit\"at Innsbruck, Technikerstrasse 25, A-6020 Innsbruck, Austria}
\author{C.~F. Roos}
\affiliation{Institut f\"ur Quantenoptik und Quanteninformation, \"Osterreichische Akademie der Wissenschaften, Otto-Hittmair-Platz 1, A-6020 Innsbruck, Austria}
\affiliation{Institut f\"ur Experimentalphysik, Universit\"at Innsbruck, Technikerstrasse 25, A-6020 Innsbruck, Austria}
\author{J.~J. Garc{\'i}a-Ripoll}
\affiliation{Instituto de F\'{\i}sica Fundamental, CSIC, Serrano 113-bis, 28006 Madrid, Spain}
\author{E. Solano}
\affiliation{Departamento de Qu\'{\i}mica F\'{\i}sica, Universidad del Pa\'{\i}s Vasco -- Euskal Herriko Unibertsitatea, Apdo.\ 644, 48080 Bilbao, Spain}
\affiliation{IKERBASQUE, Basque Foundation for Science, Alameda Urquijo 36, 48011 Bilbao, Spain}

\begin{abstract}
We propose the quantum simulation of a fermion and an antifermion field modes interacting via a bosonic field mode, and present a possible implementation with two trapped ions. This quantum platform allows for the scalable add-up of bosonic and fermionic modes, and represents an avenue towards quantum simulations of quantum field theories in perturbative and nonperturbative regimes.
\end{abstract}


\maketitle

A quantum simulator is a quantum device that can mimic the dynamics of another quantum system~\cite{Feynman82,Lloyd96}. It may represent a significant advance for studying the behavior of large quantum systems that cannot be simulated efficiently with classical computers. Some proposals include the simulation of black holes in Bose-Einstein condensates~\cite{Garay00}, particle creation in expanding universes~\cite{Alsing05,Schutzhold07}, quantum magnets~\cite{Friedenauer08}, and the emergence of relativistic physics and quantum field theories in ultracold atom lattices~\cite{Goldman09,Pachos}. In trapped-ion physics, a landmark was produced by the proposal for simulating the Dirac equation and {\it Zitterbewegung} in a single ion~\cite{Lamata07}, which was subsequently realized in the lab~\cite{Gerritsma1}. Recent advances in quantum simulations of relativistic quantum mechanics were the realization of the Klein paradox~\cite{Casanova1,Gerritsma2}, a proposal for implementing the Majorana equation and unphysical operations~\cite{Casanova2}, and the quantum simulation of two Dirac particles interacting with a classical potential~\cite{Lamata11}.

A natural step forward is to consider the constructive quantum simulation of quantized fields, in an effort to pave the way towards the implementation of quantum field theories (QFTs) in trapped-ion technologies. We point out that QFTs are significantly more involved and conceptually different from Refs.~\cite{Lamata07, Gerritsma1,Casanova1, Gerritsma2, Casanova2, Lamata11}, given that these previous works were related to relativistic quantum mechanics, i.e., a single-particle theory. QFTs are among the most successful descriptions of the physical world, and give rise to the Standard Model of elementary particles. There are different approaches to analize these theories, of which one of the most prominent is the Dyson series expansion in perturbation theory and Feynman diagrams~\cite{Peskin}. However, certain theories and parameter regimes cannot be analyzed in perturbation theory, like the strong coupling regime in which the coupling parameter is so large that perturbative methods fail. Quantum chromodynamics, for example, cannot be studied in this way for low energies or long distances. New techniques have been developed for overcoming these difficulties, as lattice gauge theory computations~\cite{Kogut}. A quantum simulator, on the other hand, could provide remarkable computational power to simulate QFTs faster than classical computers. Among all plausible implementations, trapped ions may offer one of the most versatile and powerful, due to the high degree of quantum control~\cite{LeibfriedEtAl}.

In this Letter, we propose a scalable quantum simulation of interacting bosonic and fermionic quantum field modes in trapped ions.  We consider first the case of a fermion and an antifermion field modes interacting via a bosonic field mode. This simplified model includes already some of the interesting features of QFTs: particle creation and annihilation, self interactions and dressed states, and the possibility to study nonperturbative regimes in the ultrastrong coupling (USC) or deep strong coupling (DSC) regimes~\cite{JorgeKikeDSC}. Moreover, we propose how to add progressively more fermionic and bosonic field modes with a scalable approach that may lead to the quantum simulation of a quantum field theory as quantum electrodynamics (QED). Actually, the use of lattice computations~\cite{Kogut} to make nonperturbative assertions about a continuum quantum field theory requires a careful analysis. One has to consider critical points, correctly choosing lattice spacings and coupling constants. In this sense, it might well be that theoretical developments, akin to Wilson renormalization~\cite{Wilson}, are needed so that our proposal can be used to make further predictions of quantum field theory phenomena.

We will study a model under the following assumptions: i) 1+1 dimensions; ii) scalar fermions and bosons; iii) one fermionic, one antifermionic, and one bosonic field modes. These considerations purport to make a realistic proposal with current trapped-ion experiments. Using conditions i) and ii), we begin by considering a multimode bosonic and fermionic coupled system, with Hamiltonian ($\hbar = c = 1$)
\begin{eqnarray}
\label{energy}
H =&&\int dp \,\omega (b^\dag_pb_p+d^\dag_pd_p)+\int dk \,\omega_k\, a^\dag_ka_k\nonumber\\&&+ g \int dx \  \psi^{\dag}(0,x)\psi(0,x)A(0, x),
\end{eqnarray}
where the fermionic and bosonic fields are, in
interaction picture with respect to $\int dp \,\omega (b^\dag_pb_p+d^\dag_pd_p)+\int dk \,\omega_k\, a^\dag_ka_k$,
\begin{equation}\label{ferm}
\psi(t, x)=\frac{1}{\sqrt{2\pi}} \int dp \left( b_p e^{-i \omega t} e^{i p x} + d_p^{\dag} e^{i \omega t} e^{-i p x} \right),
\end{equation}
\begin{equation}\label{bos}
A(t,x) = \frac{1}{\sqrt{2\pi}} \int dk \left( a_k e^{-i \omega_k t} e^{i k x} +  a^{\dag}_k e^{i \omega_k t} e^{-i k x} \right),
\end{equation}
and $a_k$ are the bosonic annihilation operators that follow the commutation rules $ [a_k, a^{\dag}_{k'}]= \delta(k-k') $ while $b_p$($d_p$) are fermionic(antifermionic) annihilation operators that obey the anticommutation rules $\{ b_p,b_{p'}^{\dag} \} = \delta(p-p')$ and $\{ d_p,d_{p'}^{\dag} \} =  \delta(p-p')$. Including spinors and polarization amounts to incorporating additional modes and couplings, to be addressed with a many mode extension.

In the interaction picture of Eqs.~(\ref{ferm}) and (\ref{bos}), the Hamiltonian reads,
\begin{equation}\label{energy}
H = g \int dx \  \psi^{\dag}(t, x)\psi(t, x)A(t, x).
\end{equation}
 This is a simplified version of the QED Hamiltonian~\cite{Peskin}, allowing also the description of Yukawa interactions and the coupling of fermions to the Higgs field.

\begin{figure}[t] \centering
\includegraphics[width=1\linewidth]{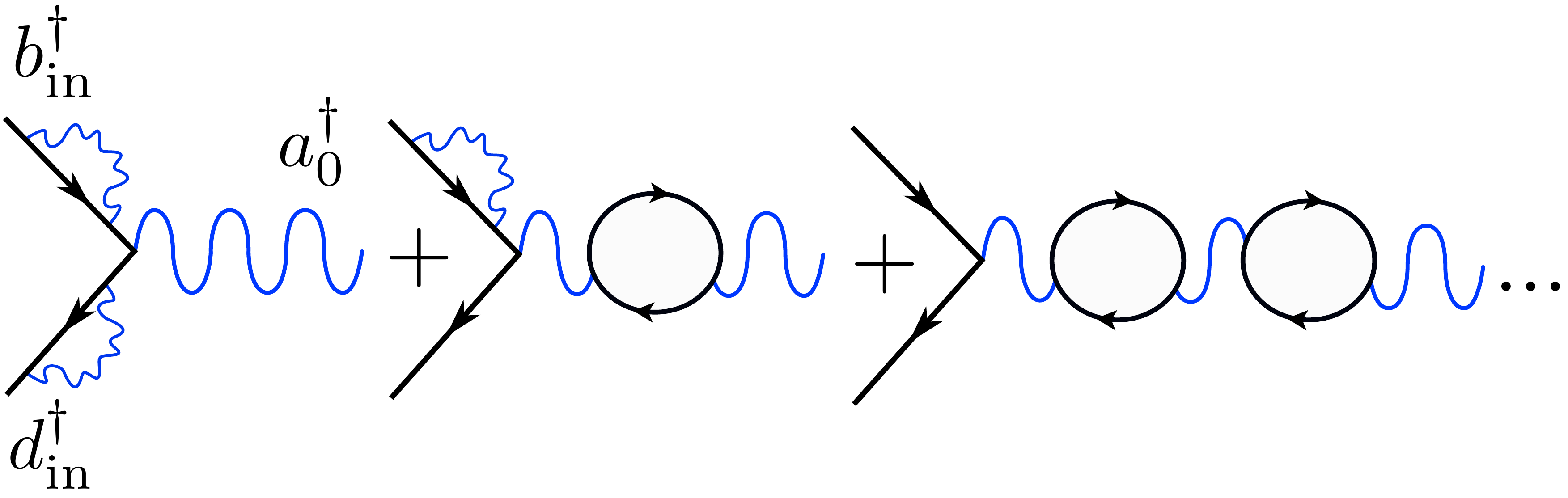}
\caption{Feynman diagrams for a number of terms in a finite-mode Dyson expansion associated with the interaction Hamiltonian in Eq. (\ref{model}). Note that loops do not imply sum over all QFT modes.}
             \label{FigFeynman}
\end{figure}

Assumption iii) amounts to introducing two {\it comoving} fermionic and antifermionic modes. They will create single-excitation incoming wave packets when applied to the vacuum, at every time $t$ in the Schr\"{o}dinger picture,
\begin{equation}
b^{\dag,{\rm Sch.}}_{\rm in}=\int dp \ {\cal{G}}_f(p_{f},p)b^{\dag}_pe^{-i\omega t},\label{fermionIncoming}
\end{equation}
\begin{equation}
d^{\dag,{\rm Sch.}}_{\rm in}=\int dp \ {\cal{G}}_{\bar{f}}(p_{\bar{f}},p)d^{\dag}_pe^{-i\omega t} .\label{antifermionIncoming}
\end{equation}
Here, ${\cal{G}}_{f,\bar{f}}(p_{f,\bar{f}},p)$ are the fermionic ($f$) and antifermionic ($\bar{f}$) envelopes of the incoming wave packets, that we consider as Gaussians centered in the average momenta, $p_{f,\bar{f}}$, with distant average initial positions. We introduce these propagating modes instead of the standard momentum eigenstates, given that the latter are delocalized over all space and are less realistic than the normalizable wave packets for describing physical particles. Although we deal with localized fermions, we consider a delocalized bosonic mode. In this way, we will have pair creation/annihilation processes and self interactions by considering just one bosonic field mode for the ease of experimental realization. We point out that our proposed quantum simulation of finite-number interacting quantized field modes includes all terms in a finite-mode Dyson expansion, instead of the standard one that takes into account all modes with a reduced number of perturbative Feynman diagrams, see Fig.~\ref{FigFeynman}. Our motivation is to have a simplified model amenable to quantum simulations with trapped ions, where the proposed theory emerges in the natural language of current experimental setups. The scalability of our approach may need specific developments of lattice theories~\cite{Kogut, Preskill2011}.

The fermionic modes $b^{\dag,{\rm Sch.}}_{\rm in}, d^{\dag,{\rm Sch.}}_{\rm in}$, are the basis for describing self-interacting dressed states, by emission and absorption of virtual bosons. They also represent the incoming states that will collide in a certain region of spacetime. These modes represent, at lowest order, the free evolution of the incoming wave packets.
Here, the pair creation and annihilation is local, i.e., it takes place when the two wave packets of fermion and antifermion overlap, as corresponds to, e.g., the QED vertex. While, on the other hand, there is an interaction at a distance between fermion and antifermion mediated by the coupling to the bosons.

Expanding the fermion field (\ref{ferm}) in terms of these modes, and dropping the remaining anticommuting modes that complete the basis (not populated at lowest order), we get, in the interaction picture,
\begin{eqnarray}
\psi(t,x) & = &  \tilde{{\cal{G}}}_f(p_{f},x,t)b_{\rm in}e^{i(p_{f}x-\omega_{f} t)}\nonumber\\&&+ \ \tilde{{\cal{G}}}_{\bar{f}}(p_{\bar{f}},x,t)d^\dag_{\rm in}e^{-i(p_{\bar{f}}x-\omega_{\bar{f}}t)},
\end{eqnarray}
where the modes $b^{\dag}_{\rm in}=\int dp \
{\cal{G}}_f(p_{f},p)b^{\dag}_p$ and $d^{\dag}_{\rm in}=\int dp \
{\cal{G}}_{\bar{f}}(p_{\bar{f}},p)d^{\dag}_p$ do not depend on time. Here,
\begin{equation}
\tilde{{\cal{G}}}_f(p_{f},x,t)=\frac{1}{\sqrt{2\pi}}\int dp \ {\cal{G}}_f(p_{f},p)e^{i[(p-p_f)x-(\omega-\omega_f) t]},
\end{equation}
\begin{equation}
\tilde{{\cal{G}}}_{\bar{f}}(p_{\bar{f}},x,t)=\frac{1}{\sqrt{2\pi}}\int dp \ {\cal{G}}^*_{\bar{f}}(p_{\bar{f}},p)e^{-i[(p-p_{\bar{f}})x-(\omega-\omega_{\bar{f}}) t]},
\end{equation}
are Fourier transforms, where we consider the ultrarelativistic limit, for simplicity, in which $\omega_{f,\bar{f}}=|p_{f,\bar{f}}|$. With this simplification, all time and space dependence in Eq. (\ref{energy}) will be encoded in $\tilde{{\cal{G}}}_f(p_{f},x,t)$, $\tilde{{\cal{G}}}_{\bar{f}}(p_{\bar{f}},x,t)$, representing Gaussian wave packets that propagate in spacetime, and in the time- and space-dependent phases that are associated to energy and momentum conservation.

The bosonic field will, in addition, be written as
\begin{equation}\label{bos1mode}
A(t,x) = a_0 e^{-i \omega_0 t} e^{i k_0 x} +  a^{\dag}_0 e^{i \omega_0 t} e^{-i k_0 x}.
\end{equation}
Consequently, the resulting interaction Hamiltonian is
\begin{equation}\label{model}
H =g\sum_{i,j =f,\bar{f} }   \ {\cal{F}}^{i,j}(p_i,p_j,k_0, t) \ \theta^{i\dag}_{p_i}\theta^j_{p_j} \ a_0  +{\rm H.c.},
\end{equation}
where $\{\theta^{i}_{p_i}\}_{i=f,\bar{f}}=b_{\rm in},d^\dag_{\rm in}$, and
\begin{eqnarray}
&& {\cal{F}}^{f,f}(p_f,p_f,k_0, t)=\left(\int dx
|\tilde{{\cal{G}}}_f(p_{f},x,t)|^2 e^{ik_0x}\right) e^{-i\omega_0 t},
 \nonumber\\&&{\cal{F}}^{\bar{f},\bar{f}}(p_{\bar{f}},p_{\bar{f}},k_0, t)=\left(\int
dx |\tilde{{\cal{G}}}_{\bar{f}}(p_{\bar{f}},x,t)|^2
e^{ik_0x}\right)e^{-i\omega_0 t},\nonumber \\&&
{\cal{F}}^{f,\bar{f}}(p_f,p_{\bar{f}},k_0, t)=\left(\int dx
\tilde{{\cal{G}}}_{f}(p_{f},x,t)^*\tilde{{\cal{G}}}_{\bar{f}}(p_{\bar{f}},x,t)\right.\nonumber
\\&&\times\left.e^{-i(p_f+p_{\bar{f}}-k_0)x}\right)e^{i(\omega_f+\omega_{\bar{f}}-\omega_0)t}.\label{funcionesAcoploF}
\end{eqnarray}
This Hamiltonian contains the {\it self-interacting} dynamics given by $|f, \bar{f}, n\rangle\leftrightarrow|f,\bar{f}, n\pm1\rangle$ ($|f, \bar{f}, n\rangle$ denotes the state with one fermion, one antifermion, and $n$ bosons), mediated by $b^{\dag}_{\rm in} b_{\rm in} a_{k_0}$, $b^{\dag}_{\rm in} b_{\rm in} a^{\dag}_{k_0}$, $d_{\rm in} d^\dag_{\rm in} a_{k_0}$ and $d_{\rm in} d^\dag_{\rm in} a^{\dag}_{k_0}$. It also includes {\it pair creation and annihilation} processes given by $|f,\bar{f}, n\rangle\leftrightarrow|0, 0, n\pm1\rangle$, mediated by $d_{\rm in}
b_{\rm in}  a^{\dag}_{k_0}$ and $b^{\dag}_{\rm in} d^{\dag}_{\rm in} a_{k_0}$ in the quasi-resonant case, as well as $d_{\rm in} b_{\rm in} a_{k_0}$ and $b^{\dag}_{\rm in} d^{\dag}_{\rm in} a^\dag_{k_0}$ in the far off-resonant case. The last kind of transitions, as well as self-interactions, are off-resonant and would be neglected in the weak coupling regime, but would be allowed in our formalism for USC/DSC regimes~\cite{JorgeKikeDSC}. In our proposed setup, all perturbative series terms are included, as shown in Fig. \ref{FigFeynman}.

\begin{figure} \centering
\includegraphics[width=\linewidth]{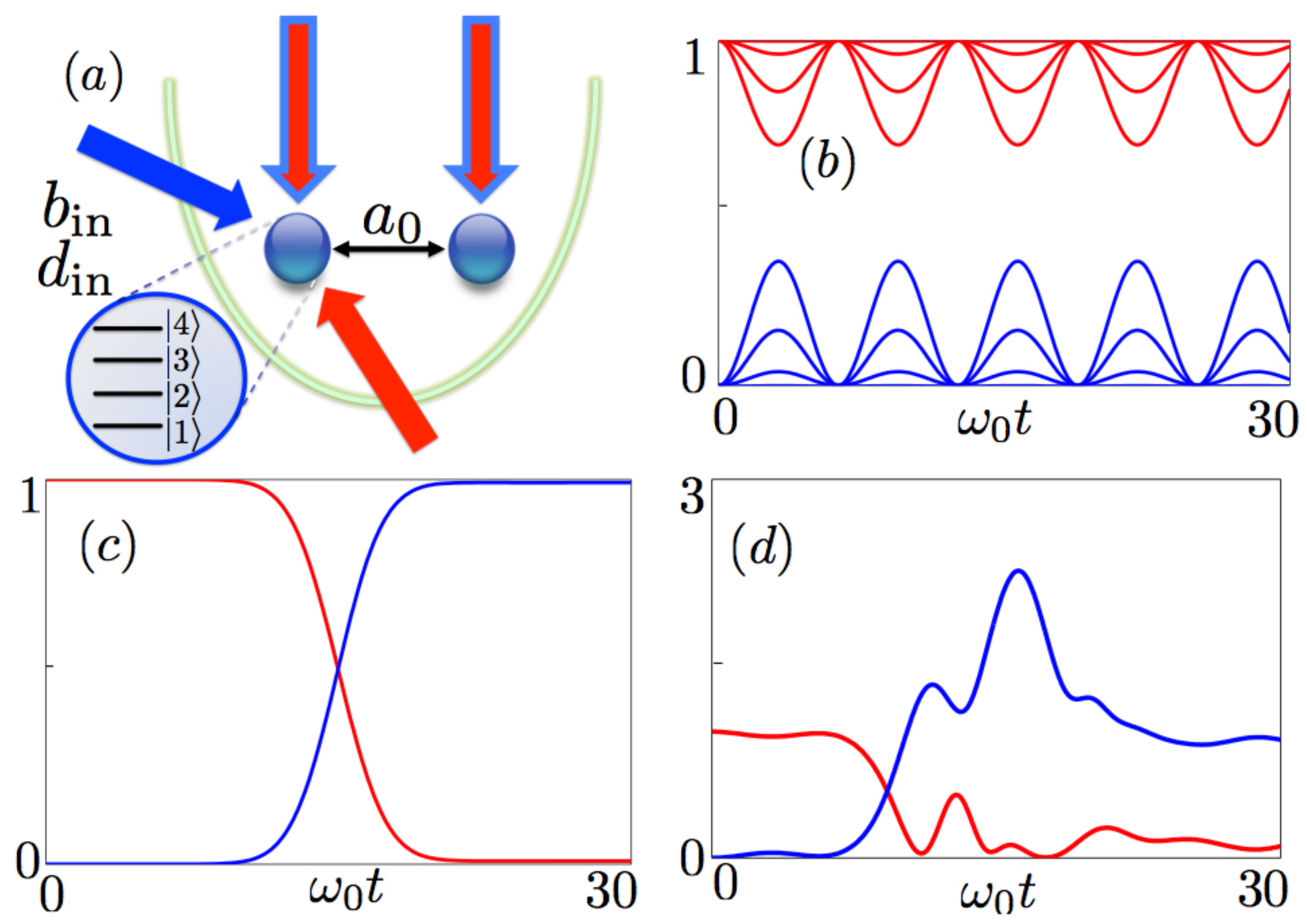}  
\caption{(a) Setup for the trapped-ion simulation. (b) $|\langle
f,0,0|\psi(t)_1\rangle|^2$ as a function of $t$ in units of $\omega_0$  (red/upper curves), where $|\psi(t)_1\rangle$ is the evolved state from $|\psi(0)_1\rangle=|f,0,0\rangle$, and average number of virtual bosons (blue/lower curves), $\langle a_0^\dag a_0\rangle$, for $g_1=0.15\omega_0, 0.1\omega_0,0.05\omega_0,0.01\omega_0$, $g_2=0$. The largest amplitudes correspond to the largest couplings.
(c) $|\langle
f,\bar{f},0|\psi(t)_2\rangle|^2$ as a function of $t$ in units of $\omega_0$  (red/upper left curve), where $|\psi(t)_2\rangle$ is the evolved state from $|\psi(0)_2\rangle=|f,\bar{f},0\rangle$, and average number of virtual bosons (blue/lower left curve), $\langle a_0^\dag a_0\rangle$, for $g_1=0.01\omega_0$, $g_2=0.21\omega_0$, $\sigma_t=3/\omega_0$, $T=30/\omega_0$, $\delta=0$.
(d) The same as (c) for $g_1=0.1\omega_0$, $g_2=\omega_0$, $\sigma_t=4/\omega_0$, $T=30/\omega_0$, $\delta=0$. }
   \label{Fig1}
\end{figure}

For practical purposes, we consider $|k_0|\ll \omega_0$, i.e., a slow massive boson. We may then approximate  ${\cal{F}}^{f,f}(p_f,p_f,k_0, t)={\cal{F}}^{\bar{f},\bar{f}}(p_{\bar{f}},p_{\bar{f}},k_0, t)=g_1\exp(-i\omega_0t)$, and ${\cal{F}}^{f,\bar{f}}(p_f,p_{\bar{f}},k_0, t)=g_2\exp[-(t-T/2)^2/(2\sigma_t^2)+i\delta t]$, where $g_2/g_1$ gives the relative strength of the pair creation with respect to the self-interaction, $\delta=\omega_f+\omega_{\hat{f}}-\omega_0$ and $T$ is the total time of the process, being $\sigma_t$ the temporal interval of the interaction region. Thus, the self-interactions are always on, while the pair creation and annihilation take place only when the fermion and antifermion wavepackets overlap, as they should. Accordingly, the Hamiltonian we aim to simulate is
\begin{eqnarray}\label{final}
&&H  =  g_1 e^{-i\omega_0 t} \left(b^{\dag}_{\rm in} b_{\rm in} a_0+d_{\rm in}d^{\dag}_{\rm in}a_0 \right) \\ & + &   g_2e^{-\frac{(t-T/2)^2}{2\sigma_t^2}}  \left[e^{ i \delta t}b^{\dag}_{\rm in} d^{\dag}_{\rm in}a_0   +  e^{-i(2\omega_0+ \delta)t}  d_{\rm in} b_{\rm in} a_0\right] + {\rm H.c.}\nonumber
\end{eqnarray}
We propose to implement this Hamiltonian dynamics in a system of two trapped ions, see Fig. \ref{Fig1}a. The bosonic mode will be encoded in the center-of-mass (COM) vibronic mode of the two-ion system. We envision to map the 4-dimensional Hilbert space associated to the fermionic/antifermionic operators onto 4 internal levels of the first ion. For this, we consider a Jordan-Wigner mapping, $b_{\rm in}^{\dag}=I\otimes\sigma^{+}$, $b_{\rm in}= I\otimes\sigma^{-}$, $d_{\rm in}^{\dag}=\sigma^{+}\otimes\sigma_z$, $d_{\rm in}=\sigma^{-}\otimes\sigma_z$,
and encode it in four internal levels of the first ion, $|1\rangle$, $|2\rangle$, $|3\rangle$, $|4\rangle$, e.g., $b^{\dag}_{\rm in} =|4\rangle\langle 3|+|2\rangle\langle 1|$,  $d^{\dag}_{\rm in} =|4\rangle\langle 2|-|3\rangle\langle 1|$, the vacuum state is state $|1\rangle$, and $|f\rangle=|2\rangle$, $|\bar{f}\rangle=-|3\rangle$, $|f,\bar{f}\rangle=-|4\rangle$. Accordingly, Hamiltonian (\ref{final}) results in
\begin{eqnarray}
\nonumber H& = & -g(t)  \bigg( |4\rangle\langle 1| \ a_0e^{ i \delta t} +  |1\rangle\langle 4| \ a^{\dag}_0e^{- i \delta t}  \bigg)\\
\nonumber &&- g(t)  \bigg( |1\rangle\langle 4| \ a_0  e^{-i(2\omega_0+ \delta )t} + |4\rangle\langle 1| \ a^{\dag}_0  e^{i(2\omega_0+ \delta )t} \bigg)\\
\nonumber &&- g_1 \bigg(|3\rangle\langle 3|  - |2\rangle\langle 2| \bigg) \bigg( a_0e^{-i\omega_0 t} + a^{\dag}_0e^{i\omega_0 t} \bigg)\\
&&+g_1 \   I \ \bigg( a_0e^{-i\omega_0 t} + a^{\dag}_0e^{i\omega_0 t} \bigg) .
\label{IonH}
\end{eqnarray}
Here, the first line corresponds to a detuned red sideband interaction between $|4\rangle$ and $|1\rangle$ with time-dependent Rabi frequency $g(t) = g_2 \exp [{- (t-T/2)^2 / 2\sigma_t^2} ]$. The second line is a detuned  blue sideband interaction, between the same levels and with the same Rabi frequency.  The third line can be developed applying  detuned red and blue sideband interactions  to $|3\rangle$ and $|2\rangle$ to produce $(|3\rangle\langle 2|- |2\rangle\langle 3|)[a_0 \exp(-i\omega_0 t) + a^{\dag}_0 \exp (i\omega_0 t) ]/i$, and a rotation of $|3\rangle$ and $|2\rangle$ with a classical field to produce the change $(|3\rangle\langle 2|- |2\rangle\langle 3|)/i \longrightarrow (|3\rangle\langle 3|  - |2\rangle\langle 2| ).$ These operations are equivalent to a rotation  from $\sigma_y$ to $\sigma_z$
in the subspace of $|3\rangle$ and $|2\rangle$ and, therefore, not affecting the other Hamiltonian terms. To implement the last line we could just periodically drive the trap, but this method is only applicable to the COM mode. The alternative scalable approach that we propose, applicable to any mode, consists of applying to the second ion detuned red and blue sidebands  to generate the term $g_1\ \sigma_x [a_0\exp(-i\omega_0 t) + a_0^{\dag}\exp(i\omega_0 t)]$ where the matrix $\sigma_x$ corresponds to the second ion internal levels. Preparing this ion into an eigenstate of $\sigma_x$, i.e  $\sigma_x |\psi\rangle = \pm |\psi\rangle$, we produce the desired interaction  $a_0\exp(-i\omega_0 t) + a_0^{\dag}\exp(i\omega_0 t)$~\cite{Gerritsma2}. Note that in the limit of large $|t|$ in Eq.~(\ref{IonH}), we could also study the spectrum properties in a straightforward manner.

We have numerically analyzed some of the features that could be simulated in the experiment: {\it i) Fermion self-interaction}--. In Fig. \ref{Fig1}b, we show the self-energy processes of a single fermion, which emits and reabsorbs bosons as it evolves. The average number of emitted bosons grows with the coupling. The effective Hamiltonian that results from projecting Eq.~(\ref{final}) onto a single initial fermion, coincides with the DSC Hamiltonian of Ref.~\cite{JorgeKikeDSC}, Eq.~(4), for qubit frequency equal to zero. Accordingly, what is observed is emission and reabsorption of the bosons at a period which goes as $2\pi/\omega_0$, where $\omega_0$ is in our case the simulated boson frequency. {\it ii) Pair creation/annihilation}--.  In Fig. \ref{Fig1}c, we show how a fermion/antifermion pair is annihilated, giving rise to a boson. {\it iii) Nonperturbative regime}--. For large $g_2$ values, $g_2\geq \omega_0$, we enter into the nonperturbative regime, in which Feynman-diagram techniques are of little help. Here, we obtain that the number of created bosons is much larger due to the nonresonant terms (see Fig. \ref{Fig1}d).  The dynamics becomes more complex and strongly dependent on the specific coupling values.

Having presented the basic three-mode model, we discuss plausible steps in an effort to approach a full-fledged quantum field theory. We envision a feasible scalable method that allows a given trapped-ion setup to add-up more fermionic and bosonic field modes. The resulting Hamiltonian, equivalent to Eq. (\ref{model}) with extension to many modes, will be
\begin{equation}\label{modelManyModes}
H =g\sum_l\sum_{i,j}   \ {\cal{F}}^{i,j}(p_i,p_j,k_l, t) \ \theta^{i\dag}_{p_i}\theta^j_{p_j} \ a_{k_l} +{\rm H.c.},
\end{equation}
with $i,j=1,...,n_f$, $n_f$ the number of fermionic/antifermionic
modes, and $l=1,...,n_b$, with $n_b$ the number of bosonic modes.
Now the functions ${\cal{F}}^{i,j}(p_i,p_j,k_l, t) $ will contain
the overlaps between the overcomplete set of Gaussians associated to
the different incoming modes, ${\cal{G}}_{i}(p_{i},p)$,
$i=1,...,n_f$, similarly to Eqs. (\ref{funcionesAcoploF}). We can consider  additional field modes by involving more ions, including internal and external degrees of freedom, aiming at an intermediate scalability~\cite{VerstraeteCirac05}. Here, it would be advantageous to make use of the transverse modes of a large ion string~\cite{Kim:2009}. These modes can be closely spaced in frequency, such that it is not necessary to generate laser frequencies for each of them, relaxing infrastructural requirements while approximating the continuum to a certain extent.  For the extension to many field modes, it is possible to consider the use of a digital simulator approach using many ions. The nonlocal spin operators could in principle be realized via recently developed stroboscopic techniques~\cite{Mueller10,Barreiro11,Lanyon11, CasanovaFermions, Mezzacapo1}. We point out that already with 10 two-level ions and 5 phonons per ion, one could perform quantum simulations of interacting quantum field modes that are beyond the reach of classical computers, that is, a Hilbert space dimension of ${10}^{10} \sim 2^{33}$.

J. C. acknowledges support from Basque Government grant BFI08.211. L. L. thanks the European Commission for a Marie Curie IEF grant.  I. L. E. is grateful to Basque Government grant IT559-10. J. J. G.-R. acknowledges funding from Spanish MICINN projects FIS2009-10061 and CAM project QUITEMAD S2009-ESP-1594. E. S. is grateful to Basque Government grant IT472-10, Spanish MICINN FIS2009-12773-C02-01, SOLID and CCQED European projects.


\begin{thebibliography}{10}

\bibitem{Feynman82}
R. P. Feynman,  Int. J. Theor. Phys. \textbf{21}, 467 (1982).

\bibitem{Lloyd96}
S. Lloyd, Science {\bf 273}, 1073 (1996).

\bibitem{Garay00}
L. J. Garay, J. R. Anglin, J. I. Cirac, and P. Zoller, Phys. Rev. Lett. {\bf 85}, 4643 (2000).

\bibitem{Alsing05}
P. M. Alsing, J. P. Dowling, and G. J. Milburn, Phys. Rev. Lett.
{\bf 94}, 220401 (2005).

\bibitem{Schutzhold07} R. Sch\"{u}tzhold et al., Phys. Rev. Lett. {\bf 99}, 201301 (2007).

\bibitem{Friedenauer08} A. Friedenauer et al., Nature Phys. {\bf 4}, 757 (2008).

\bibitem{Goldman09}
N. Goldman, A. Kubasiak, A. Bermudez, P. Gaspard, M. Lewenstein, and M. A. Martin-Delgado, Phys. Rev. Lett. {\bf 103}, 035301 (2009).

\bibitem{Pachos} J. I. Cirac, P. Maraner, and J. K. Pachos, Phys.
Rev. Lett. {\bf 105}, 190403 (2010).

\bibitem{Lamata07} L. Lamata, J. Le\'on, T. Sch\"{a}tz, and E. Solano, Phys. Rev. Lett. {\bf 98}, 253005 (2007).

\bibitem{Gerritsma1} R. Gerritsma et al., Nature {\bf 463}, 68 (2010).

\bibitem{Casanova1} J. Casanova et al., Phys. Rev. A {\bf 82}, 020101 (2010)

\bibitem{Gerritsma2} R. Gerritsma et al., Phys. Rev. Lett. {\bf 106}, 060503 (2011).

\bibitem{Casanova2} J. Casanova et al., Phys. Rev. X (in press), eprint arXiv:1102.1651 

\bibitem{Lamata11}
L. Lamata, J. Casanova, R. Gerritsma, C. F. Roos, J. J. Garc\'{\i}a-Ripoll, and E. Solano, New J. Phys. {\bf 13}, 095003 (2011).

\bibitem{Peskin} M. E. Peskin and D. V. Schroeder, Quantum Field Theory (Westview Press, 1995).

\bibitem{Kogut} J. B. Kogut, Rev. Mod. Phys. {\bf 51}, 659 (1979); J. B. Kogut and M. A. Stephanov, The Phases of Quantum Chromodynamics (Cambridge University Press, Cambridge, England, 2003).

\bibitem{LeibfriedEtAl} D. Leibfried, R. Blatt, C. Monroe, and D. Wineland, Rev. Mod. Phys. {\bf 75}, 281 (2003).

\bibitem{JorgeKikeDSC} J. Casanova et al., Phys. Rev. Lett. {\bf 105}, 263603 (2010).

\bibitem{Wilson}
Kenneth G. Wilson, Rev. Mod. Phys. {\bf 55}, 583 (1983).

\bibitem{Preskill2011} S. P. Jordan, K. S. M. Lee, and J. Preskill, eprint arXiv:1111.3633


\bibitem{VerstraeteCirac05} F. Verstraete and J. I. Cirac, J. Stat. Mech. P09012 (2005).

\bibitem{Kim:2009} K. Kim et al., Phys. Rev. Lett. {\bf 103}, 120502 (2009).

\bibitem{Mueller10} M. Mueller, K. Hammerer, Y. L. Zhou, C. F. Roos, P. Zoller, eprint arXiv:1104.2507

\bibitem{Barreiro11} J. T. Barreiro et al., Nature {\bf 470}, 486 (2011).

\bibitem{Lanyon11} B. P. Lanyon et al., Science {\bf 334}, 57 (2011).

\bibitem{CasanovaFermions} J. Casanova, A. Mezzacapo, L. Lamata, and E. Solano, eprint arXiv:1110.3730

\bibitem{Mezzacapo1} A. Mezzacapo, J. Casanova, L. Lamata, and E. Solano, eprint arXiv:1111.5603

\end{thebibliography}
\end{document}